\documentstyle[twoside,fleqn,espcrc2,epsf]{article}

\title{Domain-wall fermions with U(1) dynamical gauge fields\\
	in (4+1)-dimensions \thanks{presented by K. Nagai}}

\author{Sinya Aoki and Kei-ichi Nagai
\address{Institute of Physics,  University of Tsukuba,
         Tsukuba, Ibaraki 305, Japan}}

\begin{document}

\begin{abstract}
We carry out a numerical simulation
of a domain-wall model in (4+1) dimensions,
in the presence of a quenched U(1) dynamical gauge field
only in an extra dimension,
corresponding to the weak coupling limit 
of a (4-dimensional) physical gauge coupling.
Our numerical data suggest that
the zero mode seems absent in the symmetric phase,
so that it is difficult to construct a lattice chiral gauge theory 
in the continuum limit.
\end{abstract}

\maketitle

\section{Introduction}
Construction of chiral gauge theory is one 
of the long standing problems of lattice field theories.
Because of the fermion doubling problem,
the lattice field theory discretized naively becomes non-chiral.
Several approaches on lattice have been proposed to overcome this
difficulties, but none of them have been proven to be successful.

Recently Kaplan has proposed a domain-wall model
in order to construct lattice chiral gauge theories\cite{kaplan}.
The model consists of Wilson fermion action
in (2$n$+1) dimensions with a fermion mass term 
being the shape of a domain wall in the (extra) (2$n$+1)th dimension.
In the case of free fermions
it is shown for $0 < m_0 < 1$, where $m_0$ denotes 
the domain wall mass height, that
a massless chiral state arises as a zero mode
bound to the $2n$-dimensional domain wall 
while all doublers have large masses of the lattice cutoff scale.

In a way of introducing dynamical gauge fields
two variants of this model have been proposed:
the waveguide model\cite{waveg} and the overlap formula\cite{overl}.
However, it has been reported that the chiral zero mode disappears 
for these two variants even in the weak coupling limit, 
due to the roughness of gauge field\cite{waveg}.
In the original model
the roughness of the gauge field is replaced
with the dynamical gauge field in the extra dimension, and
in the weak coupling limit of the $2n$ dimensional coupling,
$2n$ dimensional links $U_{\mu}(x,s)$ = 1 ($\mu = 1,\cdots ,2n$)
while only extra dimensional links $U_{2n+1}(x,s)$ become dynamical.
We would like to know whether the fermionic zero modes exist 
on the domain wall or not in this limit.
This question has already been investigated 
in (2+1) dimensions\cite{aokinagai},
but the result, in particular, in the symmetric phase 
is not conclusive, due to peculiarity of the phase transition 
in the 2 dimensional U(1) spin model.
Therefore we numerically investigate this model with U(1) gauge field
in (4+1) dimensions and report the result here.

\section{Mean-field analysis}
Before numerical investigation
we estimate the effect of the dynamical gauge field in the extra
dimension using the mean-field analysis.
In our mean-field analysis all link variables in the extra dimension 
are replaced with $(x , s)$-independent constant $z$ $(0<z<1)$,
so that the fermion propagator is easily obtained\cite{aokinagai}
\footnote{Note that $G_L$ $(G_R)$ here was denoted $G_R$ $(G_L)$ in Ref.\cite{aokinagai}}
%
\begin{eqnarray}
G(p)_{s,t} = \left[ \left( -i \sum_{\mu} \gamma_{\mu} {\bar p_{\mu}} + M(z) \right) G_{R}(p) P_L \right. \nonumber \\
  + \left. \left( -i \sum_{\mu} \gamma_{\mu} {\bar p_{\mu}} + M^{\dag}(z) \right) G_L(p) P_R \right]_{s,t} \nonumber ,
\end{eqnarray}
%
where $\bar p_{\mu} \equiv \sin(p_{\mu})$ 
and $P_{R/L} = (1 \pm \gamma_5)/2$.
Corresponding fermion masses are obtained from $G_R$ and $G_L$ 
in ${\bar p} \rightarrow 0$ limit, and
we found that fermionic zero modes exist only for
$1-z < m_0 < 1$. Thus the critical value of domain wall mass
is $m_{0}^{c} = 1 - z$ and therefore no zero mode survives if $z=0$.

\section{Numerical analysis}
We investigated the (4+1) dimensional U(1) model numerically,
using a quenched approximation.
At the zero physical gauge coupling
the gauge field action of the model is reduced to
the $2n$ dimensional spin model with many copies:
\begin{equation}
S_G = \beta_s \sum_{s,x,{\hat{\mu}}} {\rm{Re}} {\rm{Tr}}
\left[ U_{D}(x,s) U^{\dag}_{D}(x+{\hat{\mu}},s) \right],
\end{equation}
where $D=2n+1$.
Therefore there exists a phase transition
which separates a broken phase from a symmetric phase.
We calculated the order parameter $v$ 
using a rotational technique\cite{aokinagai}
and found $\beta_s=0.29$ corresponds to the symmetric phase
and  $\beta_s=0.5$ to the broken phase.
We calculated the fermion propagator over 50 configurations
at $\beta_s=0.5$ and 0.29
on $L^3 \times 32\times 16$ lattices with $L=$4, 6, 8.
For the fermion field we take periodic boundary conditions
except the anti-periodic boundary condition in the 4th direction.
The errors are estimated by the single elimination jack-knife method.

At $s=0$ 
we obtained the inverse propagator $G_{R}^{-1}$ and $G_{L}^{-1}$ 
for $p_1, p_2, p_3 = 0$ as a function of $p_4$.
We obtain fermion mass squared, $m_{f}^2$, 
by extrapolating $G_{R}^{-1}$ and $G_{L}^{-1}$ linearly to $p_4 = 0$.

\subsection{Broken phase}
\begin{figure}[t]
\centerline{\epsfxsize=7.5cm \epsfbox{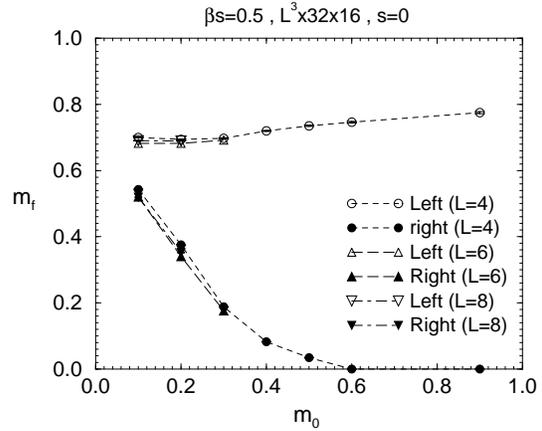}}
\vspace*{-12mm}
\caption{$m_f$ vs. $m_0$ in the broken phase}
\label{fig:broken}
\vspace*{-5mm}
\end{figure}

In Fig.\ref{fig:broken} 
we plotted $m_f$ in the broken phase ($\beta_s=0.5$)
as a function of $m_0$.
As seen from this figure, the finite size effect is small,
and the left-handed modes are always massive,
while the right-handed modes are massless if $m_0$ is larger than about $0.6$.
Therefore, we conclude that chiral zero modes can exist in the broken phase.

\subsection{Symmetric phase}
\begin{figure}[t]
\centerline{\epsfxsize=7.5cm \epsfbox{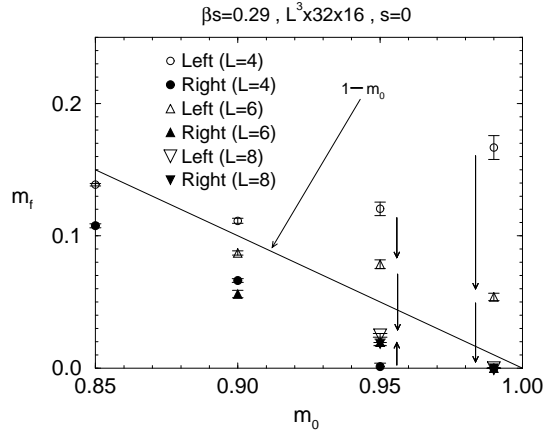}}
\vspace*{-12mm}
\caption{$m_f$ vs. $m_0$ in the symmetric phase}
\label{fig:symmetric}
\vspace*{-5mm}
\end{figure}

Let us show the fermion mass in the symmetric phase ($\beta_s=0.29$)
in Fig.\ref{fig:symmetric}.
In the smallest lattice size the chiral zero modes seem to exist.
However, for large lattices,
the mass difference between the left- and right-handed modes 
becomes smaller. This suggest that 
the fermion spectrum becomes vector-like in the infinite volume limit.
However, since the fermion mass near $m_0=1.0$ is so small,
from this data alone,
we cannot exclude a possibility that the critical mass $m_{0}^c$ is 
very close to $1.0$.

\begin{figure}[t]
\centerline{\epsfxsize=7.5cm \epsfbox{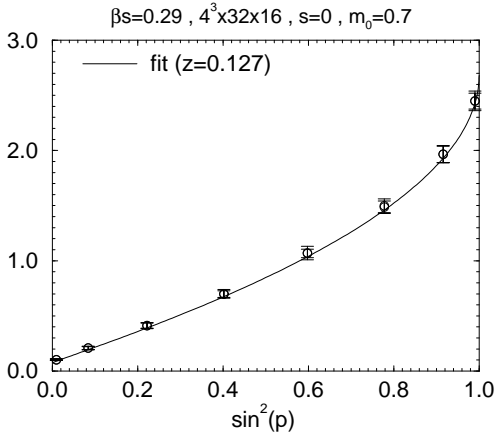}}
\vspace*{-12mm}
\caption{$G_{R}^{-1}$ vs. $\sin^2(p)$ in the symmetric phase}
\label{fig:propR}
\vspace*{-5mm}
\end{figure}

To make a definite conclusion on the absence of chiral zero modes
in the symmetric phase, we try to fit the fermion propagator 
using the form of the mean-field propagator 
with the fitting parameter $z$.
We show the quality of the fit in Fig.\ref{fig:propR}.
This figure shows that 
the fermion propagator is well described
by the mean-field propagator.

\begin{figure}[t]
\centerline{\epsfxsize=7.5cm \epsfbox{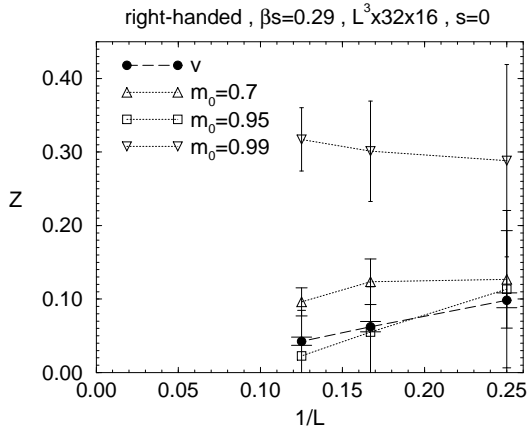}}
\vspace*{-12mm}
\caption{$z$(right-handed) vs. $1/L$ in the symmetric phase}
\label{fig:fittingR}
\vspace*{-5mm}
\end{figure}

\begin{figure}[t]
\centerline{\epsfxsize=7.5cm \epsfbox{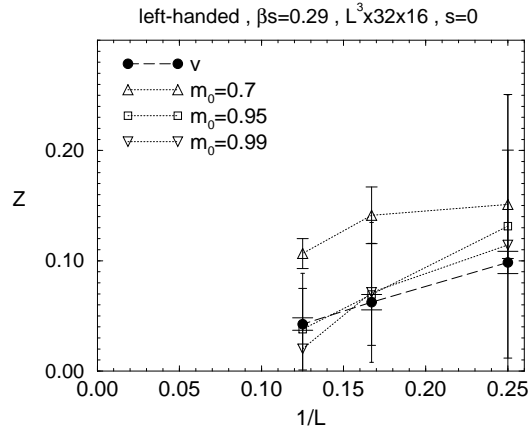}}
\vspace*{-12mm}
\caption{$z$(left-handed) vs. $1/L$ in the symmetric phase}
\label{fig:fittingL}
\vspace*{-5mm}
\end{figure}

In Fig.\ref{fig:fittingR} and Fig.\ref{fig:fittingL}
we plotted the parameter $z$ obtained the above fit 
as a function of $1/L$
The parameter $z$'s are almost independent of $m_0$ at the each $1/L$
except for the right-handed ones at $m_0=0.99$.
The solid circles represent the order parameter $v$.
The behaviors of $z$ at different $m_0$ 
are almost identical each other
and are very similar to that of $v$ 
except the right-handed ones at $m_0=0.99$.
This suggest that 
$z$ can be identified with $v$
and therefore $z$ becomes zero 
as the lattice size goes to the infinity.
If this is the case
the fermion spectrum of this model becomes vector-like
in the symmetric phase.

\section{Conclusions}
We have carried out the numerical simulations 
of the U(1) original domain-wall model
in (4+1) dimensions 
in the weak coupling limit of the 4-dimensional
coupling.

In the broken phase,
there exist chiral zero modes on the domain wall 
for $m_0 > m_{0}^c$.
The existence of the critical mass $m_{0}^c$
is predicted by the mean-field analysis.
On the other hand,
in the symmetric phase,
the analysis using the mean-field propagator suggests
that this model becomes vector-like.
We should note, however, that
the right-handed modes at  $m_0=0.99$ behaves differently
and the similar behavior was also found 
in the (2+1) dimensional model\cite{aokinagai}.
Therefore in the future,
we must investigate this point in detail, 
for example, increasing the statistics.

Besides this point
the results from both phases suggest that
this model becomes vector-like in the continuum limit,
which should be taken at the critical point 
(: the point between two phases).
Therefore,
it seems also difficult 
to construct a chiral gauge theory on lattice
via the original domain-wall model.

\end{document}